\documentclass[12pt,preprint]{aastex}

\shorttitle{Automated Classification of IRAS Sources}
\shortauthors{Gupta et al.}

\received{2001 October 17}
\begin{document}


\title{Automated Classification of 2000 Bright IRAS Sources
    }


\author{Ranjan Gupta}\affil{IUCAA, Post Bag 4, Ganeshkhind, Pune-411007, India}
\email{rag@iucaa.ernet.in}

\author{Harinder P. Singh}\affil{
Department of Physics \& Astrophysics, University of Delhi, Delhi-110007, India}
\email{hpsingh@physics.du.ac.in}

\author{K. Volk and S. Kwok}\affil{
Department of Physics \& Astronomy, University of Calgary,
 500 University Dr., \\N.W., Calgary, Alberta, Canada T2N 1N4}
\email{volk@iras.ucalgary.ca ; kwok@iras.ucalgary.ca}

\begin{abstract}

An Artificial Neural Network (ANN) scheme has been employed that
uses a supervised back-propagation algorithm to classify 2000 bright sources
from the Calgary database of IRAS (Infrared Astronomical Satellite)
spectra in the region $8\mu$m to $23\mu$m.
The database has been classified into 17 predefined
classes based on the spectral morphology.
We have been able to classify over 80 percent of the sources correctly in the
first instance. The speed and robustness of the scheme will allow us to
classify the whole of the LRS database, containing more that
50,000 sources, in the near future.
\end{abstract}


\section{Introduction}

Infrared Astronomical Satellite Low Resolution Spectrometer
(LRS) recorded spectra of some 50,000 sources in blue ($8 - 15\mu$m)
with $\lambda/\Delta\,\lambda \sim 40$ and in red ($13 - 23\mu$m)
with a resolution $\sim 20$. A total of 5425 objects with better quality
spectra were included in the Atlas of low-resolution IRAS spectra
(1986, hereafter the Atlas). Volk \& Cohen (1989a) published spectra of
356 IRAS point sources with $\rm F_{\nu} (12 \mu m) > 40 Jy$ that were not
included in the Atlas. These brighter sources were
classified into nine classes based upon the spectral morphology. Sixty
percent of the sources have silicate emission and red-continuum spectra
associated with H II region sources. No emission-line sources formed part
of the set of 356 spectra. This sample was also used to test the
classification scheme of IRAS sources based on broad-band colors.
Classifiable spectra were found for 338 of the sources in the sample.
The remaining 18 sources had either extremely noisy or incomplete spectra.
It was found that some class of sources overlapped on the color-color
diagrams and, therefore, the nature of some of the IRAS sources could
not be determined from the IRAS photometry.

Volk et al. (1991) published an additional 486 spectra belonging to sources
with $\rm 12 \mu m$ fluxes between $20$ and $\rm 40 Jy$ that were also not in the
Atlas. Classifiable spectra were found for 424 sources. The spectra were
classified into nine groups as in
Volk \& Cohen (1989a) that describe the astrophysical nature of these
sources. Kwok, Volk \& Bidelman (1997) processed 11,224 spectra (including sources
in the Atlas), corresponding to a flux limit of $\rm 7 Jy$ at $\rm 12 \mu m$.
These spectra were also classified by-eye and put into nine classes based
on the presence of emission and absorption features and on the shape of the
continuum. They identified optical counterparts of these
IRAS sources in the existing optical and infrared catalogs and listed
the optical spectral types if they were known.

It is evident that large databases like
the one referred to above require automated schemes for any analysis.
Artificial Neural Networks (ANN) have been employed extensively in several
branches of Astronomy for automated data analysis 
(Lahav \& Storrie-Lombardi, 1994).
ANN have been used previously
by the IUCAA group in three distinct areas of stellar Astronomy.
They have been applied to classify digitized optical and ultraviolet
spectra (Gulati et. al. 1994a,b; Singh et al. 1998); to compare a
set of observed spectra of F \& G dwarfs with a library of synthetic
spectra (Gulati et al. 1997a, Gupta et al. 2001); and to determine
the reddening properties of hot stars from the low-dispersion
ultraviolet spectra (Gulati et al. 1997b).

We have attempted to classify 2000 brightest sources from the Atlas
into 17 classes by means of Artificial Neural Networks.
In the next section we describe features of the 17 new classes. In Section 3,
we present details of the ANN scheme. Results are discussed in Section 4
and important conclusions of the study are presented in the last section.
Further, an appendix has been added
just before the reference section, to explain the
general ANN architecture for the benefit of the readers.

\section{Spectral Classes}

\vskip 0.5cm
The set of 2000 spectra were classified into the 17 classes by-eye by one 
of us (K.~Volk).  These classifications are assumed to from the reference 
set of ``correct'' classifications.  Such by-eye classification has an element 
of subjectivity in it, and is more apt to have problems when the spectra are 
noisy.  This set of 2000 sources are the brightest sources in the LRS atlas 
and have the least problems with noise or spectral peculiarities.  This 
makes the by-eye classification as accurate as possible despite the 
element of subjectivity.

The groups used for spectral classification are an extension of the 
nine classes described in Volk \& Cohen (1989a).  The current scheme 
introduces more classes, for three reasons:

\vskip 0.5cm
\noindent (a) The ``HII region sources'' group shows a variety of features, 
and so separate classes have been made for the more obvious feature types 
so that these classes are, one hopes, more uniform;

\vskip 0.5cm
\noindent (b) A few new classes have been created within the 10 $\mu$m 
silicate feature and 11.3 $\mu$m SiC feature groups to divide the objects 
with weaker features from those with stronger features; and 

\vskip 0.5cm
\noindent (c) Some small groups of unusual sources can be identified from 
the ``usual/unknown'' group of Volk \& Cohen (1989a), hence these were given
their own classes here such as class 14 for 21 $\mu$m feature sources.

The division between the spectra with ``stronger'' and ``weaker'' features
was made at LRS type 25 or 45 (for silicate emission sources and SiC emission
feature sources respectively) since the original LRS classes are based 
upon the feature strength.  Thus the spectra with strong features, LRS 
types 25 to 29 or 45 to 49, are separated from the spectra with weaker 
features, LRS types 21 to 24 or 41 to 44.  All objects with LRS classes 
near the dividing line were examined by-eye to make sure that the LRS 
type accurately reflects the feature strength.  In some cases this 
led to spectra being placed into the ``stronger'' or ``weaker'' feature 
groups even when the LRS type is not as expected.  In the large majority 
of cases no problem with the LRS type was found.

A short summary of the 17 classes is given in Table 1, with representative 
spectra plotted in Figure 1.  A detailed description of the different classes
and their relation to the original LRS types now follows:

\vskip 0.5cm
{\noindent\bf (0) Line emission sources}

\vskip 0.5cm
This class is for spectra with strong emission lines (line peak to continuum 
ratio of 1 or larger) and no obvious dust emission or absorption features.
These are nearly all planetary nebula sources.  A few cases of HII region 
sources that may show the 12.8 $\mu$m [NeII] line could have been included 
here, but were left in the other ``HII region'' groups since the 12.8 $\mu$m 
emission line seen in LRS spectra is sometime due to an instrumental problem 
rather than being an actual emission line, and since the possible [NeII] 
lines were relatively weak.  Comparatively few HII region spectra have strong 
enough emission lines as observed by the LRS instrument to be potentially 
included in this group.  In the original LRS classification, 
these are generally classified as types 81 to 96.

The prototype spectrum in Figure 1 shows several emission lines: 9.0 $\mu$m 
[ArIII], 12.8 $\mu$m [NeII], 15.5 $\mu$m [NeV], and 18.7 $\mu$m [SIII].  
Other objects in this class tend to have fewer emission lines than this 
example.

\vskip 0.5cm
{\noindent\bf (1) Sources with stronger 11.3 $\mu$m SiC emission features}

\vskip 0.5cm
In this group one has evolved carbon stars undergoing mass loss and showing a
strong 11.3 $\mu$m emission feature.  In the original LRS classification these 
would be type 45 to 49 sources, implying the feature peak to estimated 
continuum ratio is larger than 1.648.

\vskip 0.5cm
{\noindent\bf (2) Stellar continuum sources}

\vskip 0.5cm
Here is the class for stars without circumstellar dust shells that radiate 
significantly in the 8 to 23 $\mu$m region, including Vega, Sirius, Aldebaran, 
and Antares.  Most of these sources are of K-type to early M-type.  Only a few 
have spectral types earlier than K0.  In the original LRS classification 
these would be type 16 to 19 sources

\vskip 0.5cm
{\noindent\bf (3) Featureless sources with cooler colour temperatures}

\vskip 0.5cm
This group is as in Volk \& Cohen (1989a).  The continuum colour temperature is 
lower than expected for the photospheric emission of any normal star, 
including late M-type stars.  Aside from the unusually low colour temperature 
for a stellar photosphere there is no overt sign of any dust emission.  Many 
of these sources have optical counterparts among known Mira or irregular 
variables of M-type.  In the original LRS classification these would be type 
12 to 15 sources.  All featureless continuum sources with $\lambda$F$_\lambda$ 
ratios of less than 3 between 7.9 and 11 $\mu$m but which still have declining 
continua over the LRS wavelength range are placed in this class.

\vskip 0.5cm
{\noindent\bf (4) Sources with UIR features at 7.6/8.5/11.3/12.5 $\mu$m}

\vskip 0.5cm
This group includes sources with the UIR features, except for those with 
very steep (``red'' according to the LRS atlas definition) continua which 
form class 16.  These would ideally be type 80 sources in the original 
LRS classification, but there was considerable confusion with other types of 
objects.

As long as the spectrum signal to noise ratio is reasonable it is possible to 
objectively distinguish between UIR feature spectra and silicate absorption, 
as discussed in Volk \& Cohen (1989b).

\vskip 0.5cm
{\noindent\bf (5) Sources with 10 $\mu$m silicate absorption}

\vskip 0.5cm
This class includes objects with silicate absorption features, but we attempt 
to exclude cases of compact HII regions with foreground silicate absorption 
which have much lower continuum colour temperature than the ``normal'' 
silicate absorption feature sources such as IRAS 01304$+$6211 (see Figure 1).
If the spectral continuum is rising from 7.6 to 23 $\mu$m and there is a 
silicate absorption feature at 10 $\mu$m the spectrum is put in class 9.  
In the original LRS classification class 5 sources correspond to types 31 to 
39.  An additional refinement is that sources that are in transition between 
silicate emission and silicate absorption at 10 $\mu$m are grouped in class 13.

\vskip 0.5cm
{\noindent\bf (6) Sources with strong silicate emission at 10 $\mu$m}

\vskip 0.5cm
Spectra with a falling continuum over the LRS wavelength range and having 
a silicate emission feature with a peak greater than 1.628 times the 
continuum are in this class.  These would be types 25 to 29 in 
the original LRS classification.  Similar spectra but with weaker features 
are put into class 12.

\vskip 0.5cm
{\noindent\bf (7) Sources with low temperature ($\leq 100^\circ$K) dust 
continuum emission}

\vskip 0.5cm
In this group are mostly HII region spectra which do not show very strong 
features due to silicates or the UIR features.  The continuum must rise 
over the LRS wavelength range from 7.6 to 23 $\mu$m for a spectrum to be 
included in this class, corresponding roughly to a colour temperature of 
less than 100$^\circ$K, but generally the continuum shape cannot be 
described by a single colour temperature.  In the original LRS 
classification these would have been given some type in the 60's, 
70's, or 80's.  In a few unusual cases there can be other types 
of spectra mixed into this group.  When there are accompanying strong 
features such as the 10 $\mu$m feature in emission or absorption the 
spectra are put into classes 8, 9, or 12.

\vskip 0.5cm
{\noindent\bf (8) Sources with silicate emission and a low temperature dust 
continuum}

\vskip 0.5cm
Here are spectra with continuum shape similar to that of class 7 but with 
a clear 10 $\mu$m silicate emission feature.  In principle this corresponds 
to types 61 to 69 in the original LRS classification.  Some of the sources 
are HII regions, but others are young planetary nebulae such as Hb 12, Vy 2-2, 
and SwSt-1; others are their immediate progenitors caught in the phase 
between the asymptotic giant branch phase and the planetary nebula phase 
(``protoplanetary nebulae'' or PPNs) such as HD 161796 and IRAS 18095$+$2704. 
The spectrum of 18095$+$2704 is shown in Figure 1; this has unusually strong 
features and the continuum is not as steeply rising as is typically the 
case for spectra in this class.

\vskip 0.5cm
{\noindent\bf (9) Sources with silicate emission and a low temperature dust 
continuum}

\vskip 0.5cm
This is another class where the continuum is required to rise over the LRS 
wavelength range from 7.6 to 23 $\mu$m.  Here we include spectra showing the 
10 and 18 $\mu$m silicate features in absorption.  These are nearly all 
compact HII region sources with (presumably) foreground cold dust in the 
associate molecular cloud.  In the original LRS classification these would 
be type 71 to 79 sources.

\vskip 0.5cm
{\noindent\bf (10) Extreme carbon stars}

\vskip 0.5cm
The prototype for this group is AFGL 3068 (IRAS 23166$+$1655), which is 
known to be a carbon star with an extremely optically thick dust shell.  The 
spectra are selected for this group on the basis of having low colour 
temperatures (of order 300$^\circ$K) in the 10 $\mu$M region of the spectrum 
and of having no strong emission or absorption features.  In some cases there 
is some type of weak feature around 11 $\mu$m which might be due to the 11.3 
$\mu$M SiC feature in emission or weak absorption.  These objects are 
discussed in Volk, Kwok, \& Langill (1992).

\vskip 0.5cm
{\noindent\bf (11) Low colour temperature featureless spectra}

\vskip 0.5cm
The group of spectra have low colour temperature (about 150 to 200$^\circ$K) 
featureless continua.  They have noticeably lower colour temperatures than the 
extreme carbon star spectra see Figure 1) but peak somewhere in the LRS 
wavelength range and decline thereafter, unlike the HII region spectra in 
classes 7, 9, and 16.  A few of these sources are known to be either young 
planetary nebulae or PPNs, and they all seem to have carbon-based dust shells. 
These objects are discussed by Volk, Kwok \& Woodsworth (1993).  In the 
original LRS classification they were generally assigned type 05 or 50.

\vskip 0.5cm
{\noindent\bf (12) Sources with weaker silicate emission features}

\vskip 0.5cm
This group of spectra is the analogue of class 6, but they have feature peak 
to continuum ratios of less than 1.628.  In the original LRS atlas 
classification these would be type 21 to 24 spectra.  In various cases 
the feature was not detected at all in the original LRS atlas classification 
and these were generally classified as type 15 to 18.

\vskip 0.5cm
{\noindent\bf (13) Sources with intermediate optical depth silicate dust 
shells}

\vskip 0.5cm
Prototype spectra for this class include IRAS 19192$+$0922, IRAS 17125$-$4814, 
IRAS 15119$-$6453, and IRAS 16546$-$4047 (see Figure 1).  All these 
spectra have continua which fall between 7.6 and 23 $\mu$m.  The optical 
depth of the dust shells in these objects is such that the 10 $\mu$m feature 
is in transition between emission and absorption while the 18 $\mu$m 
feature is still in emission.  In the original LRS classification these were 
often confused with SiC emission spectra, and so given erroneous types of 41 
to 45.

\vskip 0.5cm
{\noindent\bf (14) Source with the 21 $\mu$m feature}

\vskip 0.5cm
A small number of sources are known to have a dust feature at 21 $\mu$m, 
one possible identification of which is TiC grains (von Helden et al., 2000). 
The feature was discovered in the LRS spectra.  Less than 20 such sources are 
known, the brighter of which are used to define the group here.  All sources 
in this group are carbon-rich PPNs.  A few extreme carbon stars and the 
planetary nebula IC 418 appear to have very weak 21 $\mu$m features, but these 
require the {\it Infrared Space Observatory} to detect and so they are not 
included here.

\vskip 0.5cm
{\noindent\bf (15) Sources with weaker 11.3 $\mu$m SiC emission}

\vskip 0.5cm
All spectra from carbon stars with the 11.3 $\mu$m feature but with a 
feature to continuum ratio of less and 1.628 are put in this group, 
which is closely analogous to class 1.  In the original LRS classification 
these would be of type 41 to 44.

\vskip 0.5cm
{\noindent\bf (16) Cool continuum sources with strong UIR features}

\vskip 0.5cm
Any spectrum which has a continuum that rises from 7.6 to 23 $\mu$m 
(allowing for the possibility of a strong 7.6 $\mu$m feature at the 
short wavelength end of the spectrum) and strong UIR features at 7.6 
and 11.3 $\mu$m is placed in this class.  These are HII region spectra 
with UIR features.  In the original LRS classification these would be 
of type 80 or 81.  In some cases is is possible that there is an 
underlying silicate absorption component as well as the UIR features, 
causing a very large rise at the short wavelength end of the LRS 
spectrum.  However these spectra lack any 18 $\mu$m absorption feature 
which makes the presence of strong 10 $\mu$m silicate absorption less likely.

\section{ANN Scheme}

In this analysis, we have used the multi-layer back-propagation (MBPN)
neural network scheme with supervised learning
as described in Gulati et al. (1994a,b). This scheme requires a set
of spectra to be predefined as the training set and should include
all the classes of the unknown set that the network is
supposed to classify. The algorithm trains on this
training set and subsequently
in the test phase, classifies the unknown test set into the
predefined classes.

A training set of 170 spectra having 10 sources for each training class was
set aside for training purposes.
Fig. 1 shows a representative sample of each of these 17 classes.
Each spectrum consists of 93 flux values in the range of
$8-23\mu$m. We may mention that
148 numbers of the 170 training spectra were included in the
2000 test spectra for validity checks.

The ANN scheme involved using this set of 170 input spectra with 17
assigned classes (0 to 16) to train the ANN algorithm.
We refer to these classes as the catalog classes.
The ANN configuration used was 93-16-16-17,
implying 93 data points for each of the 17 output spectral classes with
two hidden layers of 16 nodes each (a detailed explanation on the
ANN architecture is given in the Appendix).
The training session involved an iterative procedure with the
network weights getting modified at each iteration.
During an iteration, the
computed output and the desired output were compared and the resultant error
was then utilized to modify the network weights for the next iteration
using a back propagation algorithm. The learning or training was stopped
once the error was minimized to a predefined level and the network weights
were considered to be frozen (Gulati et al. 1994a).
A learning curve for 10000 iterations is given in Fig. 2 (left panel).

Subsequent to the above training session, the test session 
uses the frozen weights determined above to perform the classification of the
2000 test set of spectra into 17 catalog classes.

\section{Results and Discussion}

A general picture of the result of our ANN
classification scheme is given in Fig. 2 (right panel)
in the form of a histogram. A total of 1,618
spectra have been classified correctly out of the total sample
of 2000 source spectra indicating a success rate of 80\% at the
first instance. Table 2 provides a summary of the ANN classification
of the 2000 test sources into the 17 groups or classes. Table 3
lists 382 of the misclassified sources with their catalog class
and the corresponding ANN class.
In the following, we describe some finer details of the
classification accuracy for individual classes 0 to 16.

Out of a total 23 spectra for class 0, 19 were classified
correctly. From the four that were misclassified, one each was classified
as class 7, 8, 13, and 16. Table 3 lists the 4 spectra of
class 0 that were misclassified. Source 22036$+$5306 has been put in class
16 by the ANN probably because of the strong feature at around $\rm 8 \mu m$,
typical of class 16.

Out of a total 136 spectra for class 1, 113 were classified correctly.
From the 23 that were misclassified, 11 were misclassified into class 13;
6 were misclassified into class 15, 3 were misclassified into class
4, 2 were misclassified into class 6 and 1 was misclassified into class 14.
The misclassified sources of class 1 are listed in Table 3. Out of these,
18 spectra seem to belong to the class assigned by the
ANN and not the catalog class, specially
those put into classes 13 and 15.
ANN Classification accuracy of class 2 has the best agreement with
the catalog class. Out of 89 spectra of class 2, 88 spectra
were classified correctly while 1 spectrum was misclassified into class 3.

Of the 94 spectra for class 3, 85 were classified correctly.
Of the 9 that were misclassified, 5 were misclassified into class 2
and 1 each into classes 1, 12, 13 and 15. 
At least 1 source (20056$+$1834) has a spectrum which does not seem to belong to
class 3 and has been put in class 13 by the ANN.

Of a total of 87 sources of class 4, 34 were correctly classified
while 53 were misclassified; 22 into class 5, 11 into class 16, 10 into
class 13, 4 into class 15, 2 each into lass 9 and 14 and 1 each into
class 2 and 10. 
As mentioned in Sec. 2, class 4 corresponds to type 80 sources in the
original LRS classification in which there was a lot of confusion with
the other types of sources. ANN put 22 sources into class 5, which has
sources with $\rm 10 \mu m$ silicate absorption feature. 
We note that in the training sample of class 4, there are
at least two sources (16367$-$4701 \& 19327$+$3024) which might be
resulting  in contamination of this class with class 5 and hence less
than efficient training for class 4.

Out of a total of 103 spectra for class 5, 88 were correctly
classified. 7 spectra were misclassified into class 16, 4 into class 4,
2 into class 9 and 1 each into class 13 and 14. ANN has correctly
picked up at least 3 spectra from the catalog that should be put in
class 4 (Table 3).

Class 6 (strong $10\mu$m feature) has the largest sample
of 735 sources. Out of these 679 were classified correctly and
33 were misinterpreted as class 12 (weak $10\mu$m feature). 
Similarly, class 12 has  198 spectra out of which 15 were misclassified
as class 6, while 27 were misclassified into class 3.
As discussed later, if
one were to treat a weak and a strong $10\mu$m feature sources as
belonging to a single class, percentage of correctly classified
sources will be in excess of 90\% for these classes.

For class 7, out of the 53 sources, 45 have been classified correctly. 3
sources were incorrectly classified into class 0 while
2 sources each were incorrectly classified into classes 9 and 11.
Class 8 has 13 sources out of which only one has been wrongly classified
into class 0.

Class 9 has 40 sources in all. 26 have been correctly classified, 4 each
have been misclassified into classes 7 and 14 and 3 each into classes
5 and 16. Class 10 has 44 sources in all and 38 have been correctly
classified but 3 have been wrongly classified to class 13, 2 into class 6 
and 1 into class 1. Class 11 has 12 sources out of which 8 were correctly
classified. 3 sources were incorrectly classified into class 14 and 1 into
class 0.

As mentioned in Section 2, in the original
LRS classification, class 13 spectra were often confused with
SiC feature sources. In the ANN classification also, out of the 200
class 13 sources, 104 were wrongly classified (85 into class 6)
while 96 were correctly classified. The wrongly classified
sources have been listed in Table 3 for the purpose of review by the
human classifiers.

Class 14 has 3 sources out of which one was misclassified into
class 10. Class 15 had 140 sources out of which 111
have been correctly classified. But 18 have been classified into
class 1 and 3 each into classes 13 and 4; 2 each into class 2 and 4, 
and 1 into class 3. Lastly, 25 sources out of a total of 30
of class 16 were classified correctly. 2 sources were wrongly classified into
class 7 while 1 each were classified into classes 0, 1 and 5.

If we assume classes 6 and 12 to be the same,
(class 12 trained as class 6)
and redo the classification exercise, we obtain the overall classification
histogram as shown in Fig. 3. A total of 1,675 spectra have now been
classified correctly which is about 84\% of the total test data set.
This process also improved the classification of the class 6 sources which
is shown in Fig. 5.

We split off class 12 from class 6 because it improved the classification
of silicate emission objects. With the full range of silicate emission
objects in one class there was more dispersion of values and the ANN system had
more chance of misclassification. However, it does not matter that sources
are misclassified from class 6 to class 12 or class 12 to class 6
because the dividing line between these two groups is not sharp an
there is a continuous range of silicate feature strengths, and some objects
must fall at the boundary.

Fig. 6 shows the improvement in the
class 13 classification, namely earlier 85 sources of this class were wrongly
classified as class 6 which has now reduced to 50 wrong classifications.
This process also improved the classification of class 4
as seen in Fig. 4. There were such improvements noticed in general and
Fig. 3 is really a combined effect of all these.

\section{Conclusions}

We have demonstrated in this paper the application of ANN scheme to a large
database of 2000 IRAS spectra and have been able to correctly classify
more than 80\% of the data-set. The misclassified spectra
were looked in detail and most of them could have been wrongly catalogued
or had features which would have been confusing for even a human classifier.

We stress here that the speed and robustness of this scheme can be very
useful for classifying the whole of LRS database containing
over 50,000 sources.

\acknowledgements

RG and HPS are grateful to SK and KV for kind hospitality
while on a visit to Calgary. We thank the two anonymous referees
for very useful comments.

\newpage

\appendix

\section{ANN architecture}

In a feedforward neural network there are several inputs, a few 
hidden layers, and several outputs (Bailer-Jones et. al., 2002).  
See the Figure 7 for a block view of this
architecture. Each node in the input layer
holds a value, $x_i$. In our example application, the input vector,
$(x_1,x_2,\ldots,x_i,\ldots)$, is the spectrum with 93 flux values, 
and the output vector, $(y_1,y_2,\ldots,y_l,\ldots)$, has 17 nodes.
Each of the input nodes connects to every node in the next layer
of nodes, the first `hidden' layer, and each of these connections has
a weight, $w_{i,j}$, associated with it.  A node in the hidden layer
forms a weighted sum of its inputs, and passes this through a {\it
nonlinear transfer function}, such that the output from the $j^{th}$
hidden node is

$$ p_j = \tanh \left( \sum_i w_{i,j} x_i \right) \ \ . \eqno(A.1) $$

\noindent

These values are passed to a second hidden layer which performs
a similar processing, the output from that layer being the vector $\bf{q}$

$$ q_k = \tanh \left( \sum_j w_{j,k} p_j \right) \ \ . \eqno(A.2) $$

\noindent
The output layer then performs a simple sum of its inputs, so that the
network output, $y_l$, is

$$ y_l =  \sum_k w_{k,l} q_k \ \ . \eqno(A.3) $$

\noindent
The tanh function in the hidden layers provides the nonlinear
capability of the network.  Other nonlinear functions are possible;
the sigmoidal function ($1/(1-\exp[-\sum wx])$) is used here. Both
functions map an infinite possible input range onto a finite output
range, $-1$ to $+1$ in the case of tanh. This imitates the transfer
function of neurons.

\clearpage
\begin{table}
\begin{center}
\caption{A short description of the 17 ANN training classes}
\begin{tabular}{llc}
\tableline\tableline
Source type & Description & ANN Class \\
\tableline
Stellar & Stellar photospheric spectra; B-type to early M-type & 2\\
\cline{2-3}
& Lower temp. stellar continuum spectra with no & 3\\
& features: mid to late M-type, little circumstellar dust& \\
\tableline
Carbon stars & Strong 11.3 $\rm \mu m$ emission & 1\\ \cline{2-3}
& Lower temp. continuum ($\rm \sim 250^{\circ}K$),& 10\\
& e.g., AFGL 3068 \&  no strong features &\\ \cline{2-3}
& Weaker 11.3 $\rm \mu m$ emission & 15\\
\tableline
Oxygen-rich AGB stars & Silicate absorption on an intermediate temp.& 5\\
& continuum ($\rm 400-200^{\circ}K)$ &\\ \cline{2-3}
& Stronger 10 $\rm \mu m$ emission on a & 6\\
& higher temp. continuum ($\rm \geq 600^{\circ}K$) &\\ \cline{2-3}
& Weaker 10 $\rm \mu m$ emission on a & 12\\
& higher temp. continuum ($\rm \geq 600^{\circ}K$) &\\ \cline{2-3}
& 10 $\rm \mu m$ features in transition from& 13\\
& emission to absorption &\\
\tableline
Planetary nebulae & Emission lines (PN)& 0\\ \cline{2-3}
(PN) \& post-AGB & Silicate emission on a low temp. continuum& 8\\
sources (PPN) & (oxygen-rich PN \& PPN) & \\ \cline{2-3}
& Lower temp. ($\rm \sim 150^{\circ}K$) featureless& 11\\
& (carbon-rich PN and PPN) &\\ \cline{2-3}
& 21 $\rm \mu m$ emission feature (carbon-rich PPN)& 14\\
\tableline
ISM spectra: & UIR or PAH features on a flat continuum& 4\\
mostly HII regions; & (HII regions, galaxies, a few PN \& PPN) &\\ \cline{2-3}
includes a few PN, & Low temp. continuum ($\rm \leq 100^{\circ}K$)& 7\\
PPN, \& galaxies & without strong features &\\ \cline{2-3}
& 10 $\rm \mu m$ absorption on a low temp. continuum& 9\\
& ($\rm \leq 100^{\circ}K)$, includes a few PPN &\\ \cline{2-3}
& UIR or PAH features on a low temp. continuum & 16\\
& ($\rm \leq 100^{\circ}K)$, includes a few PN &\\
\tableline
\end{tabular}
\end{center}
\end{table}
\clearpage
\begin{table}
\begin{center}
\caption{2000 Test sources classified by ANN into 17 catalog
classes.\label{tbl-2}}
\begin{tabular}{ccccccccccccccccccc}
\tableline\tableline
Catalog&
\multicolumn{17}{c}{ANN Class}&Total\\
\cline{2-18}
Class & 0 & 1&2&3&4&5&6&7&8&9&10&11&12&13&14&15&16&\\
\tableline
0&19&...&...&...&...&...&...&1&1&...&...&...&...&1&...&...&1&23\\
1&...&113&...&...&3&...&2&...&...&...&...&...&...&11&1&6&...&136\\
2&...&...&88&1&...&...&...&...&...&...&...&...&...&...&...&...&...&89\\
3&...&1&5&85&...&...&...&...&...&...&...&...&1&1&...&1&...&94\\
4&...&...&1&...&34&22&...&...&...&2&1&...&...&10&2&4&11&87\\
5&...&...&...&...&4&88&...&...&...&2&...&...&...&1&1&...&7&103\\
6&...&1&...&...&...&...&679&...&8&...&9&...&33&5&...&...&...&735\\
7&3&...&...&...&...&...&...&45&...&2&...&2&...&...&...&...&1&53\\
8&1&...&...&...&...&...&...&...&12&...&...&...&...&...&...&...&...&13\\
9&...&...&...&...&...&3&...&4&...&26&...&...&...&...&4&...&3&40\\
10&...&1&...&...&...&...&2&...&...&...&38&...&...&3&...&...&...&44\\
11&1&...&...&...&...&...&...&...&...&...&...&8&...&...&3&...&...&12\\
12&...&6&...&27&...&...&15&...&...&...&...&...&149&1&...&...&...&198\\
13&...&...&...&1&3&4&85&...&3&...&8&...&...&96&...&...&...&200\\
14&...&...&...&...&...&...&...&...&...&...&1&...&...&...&2&...&...&3\\
15&...&18&2&1&3&...&...&...&...&...&...&...&...&3&2&111&...&140\\
16&1&1&...&...&...&1&...&2&...&...&...&...&...&...&...&...&25&30\\
\tableline
\end{tabular}
\end{center}
\end{table}

\clearpage

\begin{deluxetable}{ccccccccc}
\tablecolumns{9}
\tablewidth{0pc}
\tablecaption{List of LRS sources misclassified by ANN}
\tablehead{
\colhead{IRAS} & \colhead{Catalog} & \colhead{ANN} & \colhead{IRAS} & \colhead{Catalog} & \colhead{ANN} & \colhead{IRAS} & \colhead{Catalog} & \colhead{ANN}\\
\colhead{designation} & \colhead{class} & \colhead{class} & \colhead{designation} & \colhead{class} & \colhead{class} & \colhead{designation} & \colhead{class} & \colhead{class}}
\startdata
04395$+$3601 & 0 & 13     & 15380$-$6545 & 3 & 12   & 14198$-$6115 & 4 & 5 \\
17069$-$4149 & 0 & 7      & 17282$-$5102 & 3 & 2    & 14206$-$6151 & 4 & 16 \\
21014$-$1133 & 0 & 8      & 17504$-$0234 & 3 & 2    & 15246$-$5612 & 4 & 16 \\
22036$+$5306 & 0 & 16     & 19369$+$2823 & 3 & 1    & 15535$-$5328 & 4 & 5 \\
03293$+$6038 & 1 & 15     & 20056$+$1834 & 3 & 13   & 16041$-$4912 & 4 & 5 \\
04340$+$4623 & 1 & 13     & 22476$+$4047 & 3 & 2    & 16204$-$4717 & 4 & 5 \\
05405$+$3240 & 1 & 15     & 23092$+$5236 & 3 & 15   & 16225$-$4844 & 4 & 14 \\
08011$-$3627 & 1 & 6      & 23095$+$5925 & 3 & 2    & 16331$-$4637 & 4 & 16 \\
08086$-$3905 & 1 & 13     & 00450$-$2533 & 4 & 16   & 16467$-$4255 & 4 & 13 \\
08534$-$5055 & 1 & 13     & 01056$+$6251 & 4 & 5    & 16573$-$4619 & 4 & 5 \\
08544$-$4431 & 1 & 13     & 02401$-$0013 & 4 & 10   & 17076$-$4702 & 4 & 13 \\
10098$-$5742 & 1 & 14     & 04064$+$5052 & 4 & 13   & 17324$-$3152 & 4 & 16 \\
12195$-$6830 & 1 & 13     & 05044$-$0325 & 4 & 5    & 17355$-$3241 & 4 & 2 \\
13045$-$6404 & 1 & 13     & 05345$+$3157 & 4 & 5    & 17486$-$2345 & 4 & 5 \\
13064$-$6433 & 1 & 13     & 06114$+$1745 & 4 & 16   & 18077$-$2614 & 4 & 5 \\
15084$-$5702 & 1 & 13     & 06319$-$0501 & 4 & 5    & 18092$-$1742 & 4 & 9 \\
15261$-$5702 & 1 & 15     & 07013$-$1128 & 4 & 13   & 18162$-$0246 & 4 & 13 \\
15469$-$5311 & 1 & 4      & 07017$-$1114 & 4 & 5    & 18162$-$1612 & 4 & 16 \\
16093$-$4808 & 1 & 15     & 08247$-$4223 & 4 & 5    & 18254$-$1149 & 4 & 5 \\
16192$-$4900 & 1 & 15     & 08438$-$4340 & 4 & 9    & 18310$-$2834 & 4 & 5 \\
16455$-$4349 & 1 & 4      & 09014$-$4736 & 4 & 13   & 18312$-$1209 & 4 & 5 \\
17199$-$3512 & 1 & 4      & 10105$-$5719 & 4 & 16   & 18320$-$0352 & 4 & 15 \\
17289$-$3106 & 1 & 13     & 10267$-$5658 & 4 & 13   & 18357$-$0604 & 4 & 5 \\
18301$-$0656 & 1 & 6      & 10366$-$5931 & 4 & 13   & 18463$-$0052 & 4 & 16 \\
18367$-$0452 & 1 & 13     & 11356$-$6144 & 4 & 13   & 19097$+$0847 & 4 & 16 \\
18551$+$0323 & 1 & 15     & 11418$-$6706 & 4 & 13   & 19156$-$0935 & 4 & 15 \\
21223$+$5114 & 1 & 13     & 12421$-$6217 & 4 & 15   & 19193$+$1504 & 4 & 5 \\
16063$-$4906 & 2 & 3      & 13065$-$6354 & 4 & 5    & 19343$+$2026 & 4 & 5 \\
00192$-$2020 & 3 & 2      & 14092$-$6506 & 4 & 5    & 21282$+$5050 & 4 & 14 \\
21345$+$5410 & 4 & 15     & 05390$+$1448 & 6 & 12   & 19039$+$0809 & 6 & 12 \\
22308$+$5812 & 4 & 16     & 06434$-$3628 & 6 & 12   & 19089$+$1542 & 6 & 8 \\
23541$+$7031 & 4 & 5      & 06546$-$2353 & 6 & 12   & 19231$-$2717 & 6 & 13 \\
09199$-$5447 & 5 & 14     & 08189$+$0507 & 6 & 12   & 19291$+$2012 & 6 & 10 \\
17317$-$3331 & 5 & 16     & 10174$-$5704 & 6 & 8    & 19354$+$5005 & 6 & 12 \\
17443$-$2949 & 5 & 16     & 10360$-$5633 & 6 & 12   & 19420$+$3318 & 6 & 10 \\
18168$-$1520 & 5 & 13     & 11113$-$5949 & 6 & 12   & 20004$+$2955 & 6 & 8 \\
18257$-$1000 & 5 & 16     & 11492$-$6052 & 6 & 12   & 20273$+$3932 & 6 & 1 \\
18316$-$0746 & 5 & 16     & 12188$-$6246 & 6 & 10   & 20526$-$5431 & 6 & 12 \\
18407$-$0619 & 5 & 4      & 12230$-$5943 & 6 & 12   & 21044$-$1637 & 6 & 12 \\
18491$-$0207 & 5 & 16     & 14297$-$6010 & 6 & 10   & 21389$+$5405 & 6 & 12 \\
19065$+$0832 & 5 & 16     & 16235$+$1900 & 6 & 12   & 21419$+$5832 & 6 & 12 \\
19352$+$2030 & 5 & 9      & 16275$-$2638 & 6 & 13   & 22000$+$5643 & 6 & 12 \\
20110$+$3321 & 5 & 9      & 16316$-$5026 & 6 & 12   & 22196$-$4612 & 6 & 12 \\
20187$+$4111 & 5 & 4      & 16367$-$2046 & 6 & 12   & 22480$+$6002 & 6 & 8 \\
20318$+$3829 & 5 & 4      & 16434$-$4545 & 6 & 10   & 23416$+$6130 & 6 & 8 \\
20547$+$0247 & 5 & 4      & 17189$-$6501 & 6 & 10   & 05305$+$3029 & 7 & 16 \\
23151$+$5912 & 5 & 16     & 17237$-$3102 & 6 & 12   & 09032$-$3953 & 7 & 11 \\
00001$+$4826 & 6 & 12     & 17269$-$2625 & 6 & 12   & 11143$-$6113 & 7 & 0 \\
01010$+$7434 & 6 & 12     & 17450$-$2724 & 6 & 10   & 14394$-$6004 & 7 & 9 \\
01251$+$1626 & 6 & 12     & 17515$-$2407 & 6 & 13   & 16268$-$4556 & 7 & 0 \\
01597$+$5459 & 6 & 12     & 17538$-$3728 & 6 & 12   & 16342$-$3814 & 7 & 9 \\
02360$+$5922 & 6 & 12     & 18038$-$1614 & 6 & 13   & 17078$-$3927 & 7 & 0 \\
03489$-$0131 & 6 & 12     & 18207$-$1029 & 6 & 10   & 20028$+$3910 & 7 & 11 \\
03503$+$6925 & 6 & 12     & 18243$+$0352 & 6 & 12   & 21078$+$5211 & 8 & 0 \\
04137$+$3114 & 6 & 12     & 18303$-$0519 & 6 & 13   & 06053$-$0622 & 9 & 7 \\
04188$+$2819 & 6 & 8      & 18309$-$6955 & 6 & 12   & 07399$-$1435 & 9 & 7 \\
04292$+$3100 & 6 & 12     & 18363$-$0523 & 6 & 8    & 10019$-$5712 & 9 & 5 \\
04525$+$3028 & 6 & 8      & 18518$+$0358 & 6 & 10   & 10460$-$5811 & 9 & 7 \\
13416$-$6243 & 9 & 14     & 07418$-$2850 & 12 & 6   & 18586$-$1249 & 12 & 3 \\
16313$-$4840 & 9 & 14     & 07434$-$3750 & 12 & 6   & 19055$+$0613 & 12 & 3 \\
18032$-$2032 & 9 & 16     & 09411$-$5933 & 12 & 1   & 19111$+$2555 & 12 & 1 \\
18100$-$1915 & 9 & 5      & 09508$-$4345 & 12 & 6   & 19328$+$3039 & 12 & 6 \\
18316$-$0602 & 9 & 16     & 09564$-$5837 & 12 & 3   & 19510$-$5919 & 12 & 3 \\
18317$-$0757 & 9 & 7      & 10383$-$7741 & 12 & 3   & 19585$+$5200 & 12 & 3 \\
18379$-$0500 & 9 & 5      & 11202$-$5305 & 12 & 13  & 20111$-$4708 & 12 & 3 \\
18566$+$0408 & 9 & 16     & 13248$-$7851 & 12 & 3   & 20416$+$1903 & 12 & 3 \\
19310$+$1745 & 9 & 14     & 13548$-$3049 & 12 & 1   & 20438$-$0415 & 12 & 1 \\
19566$+$3423 & 9 & 14     & 14129$-$5940 & 12 & 6   & 20507$+$2310 & 12 & 6 \\
02152$+$2822 & 10 & 1     & 14234$-$5359 & 12 & 3   & 23212$+$3927 & 12 & 1 \\
06505$-$0450 & 10 & 13    & 15410$-$0133 & 12 & 6   & 05073$+$5248 & 13 & 6 \\
13522$-$5619 & 10 & 13    & 15492$+$4837 & 12 & 3   & 06259$-$1301 & 13 & 6 \\
14318$-$5937 & 10 & 6     & 16052$-$2339 & 12 & 3   & 06308$+$0402 & 13 & 5 \\
16279$-$4709 & 10 & 13    & 16306$+$7223 & 12 & 3   & 06364$+$0846 & 13 & 8 \\
17056$-$3959 & 10 & 6     & 16328$-$4656 & 12 & 1   & 06491$-$0654 & 13 & 6 \\
05573$+$3156 & 11 & 14    & 16334$-$3107 & 12 & 3   & 07180$-$1314 & 13 & 6 \\
09370$-$4826 & 11 & 14    & 16383$-$1952 & 12 & 6   & 07376$-$2827 & 13 & 6 \\
13308$-$6209 & 11 & 0     & 16387$-$2700 & 12 & 3   & 08357$-$1013 & 13 & 6 \\
19500$-$1709 & 11 & 14    & 16438$-$1133 & 12 & 6   & 09354$-$5627 & 13 & 10 \\
00007$+$5524 & 12 & 6     & 16494$-$1252 & 12 & 6   & 10028$-$5825 & 13 & 8 \\
00245$-$0652 & 12 & 6     & 16534$-$3030 & 12 & 3   & 10077$-$5304 & 13 & 6 \\
01150$+$5732 & 12 & 3     & 17201$-$4613 & 12 & 3   & 10277$-$5730 & 13 & 4 \\
03082$+$1436 & 12 & 3     & 17297$+$1747 & 12 & 3   & 10287$-$5733 & 13 & 6 \\
05132$+$5331 & 12 & 6     & 17318$-$2342 & 12 & 6   & 10323$-$4611 & 13 & 6 \\
05450$-$3142 & 12 & 6     & 17505$-$7021 & 12 & 3   & 10481$-$6930 & 13 & 6 \\
06210$+$4918 & 12 & 3     & 17578$-$1700 & 12 & 3   & 11214$-$6448 & 13 & 6 \\
06261$+$1637 & 12 & 3     & 18347$-$0241 & 12 & 3   & 11296$-$4431 & 13 & 6 \\
07091$-$2902 & 12 & 3     & 18378$-$3731 & 12 & 3   & 11467$-$6234 & 13 & 6 \\
11528$-$5902 & 13 & 6     & 16320$-$4419 & 13 & 10  & 18061$-$1743 & 13 & 6 \\
11575$-$7754 & 13 & 6     & 16335$-$4707 & 13 & 6   & 18083$-$2630 & 13 & 6 \\
12043$-$6225 & 13 & 6     & 16350$-$4754 & 13 & 6   & 18085$-$1832 & 13 & 6 \\
12222$-$4652 & 13 & 6     & 16365$-$4717 & 13 & 6   & 18089$-$2952 & 13 & 3 \\
12384$-$4536 & 13 & 6     & 16399$-$3548 & 13 & 10  & 18222$-$1544A & 13 & 6 \\
13243$-$6159 & 13 & 6     & 16409$-$5128 & 13 & 6   & 18373$-$0021 & 13 & 6 \\
13366$-$6222 & 13 & 6     & 16414$-$4941 & 13 & 6   & 18373$-$0922 & 13 & 6 \\
13517$-$6515 & 13 & 6     & 16490$-$4618 & 13 & 6   & 18381$+$0020 & 13 & 6 \\
13527$-$6117 & 13 & 10    & 16538$-$4135 & 13 & 6   & 18409$+$0431 & 13 & 6 \\
14273$-$6153 & 13 & 6     & 16567$-$4659 & 13 & 6   & 18471$-$0259 & 13 & 6 \\
14591$-$4438 & 13 & 6     & 16580$-$4424 & 13 & 6   & 18530$+$0817 & 13 & 6 \\
15008$-$5808 & 13 & 6     & 16598$-$4117 & 13 & 6   & 18588$-$1915 & 13 & 6 \\
15044$-$5822 & 13 & 6     & 17030$-$4246 & 13 & 10  & 19007$-$3826 & 13 & 6 \\
15054$-$5458 & 13 & 10    & 17109$-$3243 & 13 & 6   & 19043$+$1009 & 13 & 6 \\
15152$-$6241 & 13 & 6     & 17132$-$5003 & 13 & 6   & 19135$+$0931 & 13 & 6 \\
15163$-$5525 & 13 & 8     & 17186$-$4208 & 13 & 6   & 19171$+$1119 & 13 & 5 \\
15174$-$4821 & 13 & 6     & 17239$-$2812 & 13 & 6   & 19229$+$1708 & 13 & 6 \\
15236$-$5556 & 13 & 6     & 17362$-$3322 & 13 & 6   & 19244$+$1809 & 13 & 6 \\
15408$-$5413 & 13 & 4     & 17368$-$3000 & 13 & 6   & 19493$+$2905 & 13 & 10 \\
15500$-$5135 & 13 & 6     & 17382$-$1704 & 13 & 6   & 19520$+$2759 & 13 & 4 \\
15527$-$6041 & 13 & 6     & 17401$-$5730 & 13 & 6   & 20095$+$2726 & 13 & 6 \\
15532$-$4802 & 13 & 6     & 17479$-$2927 & 13 & 6   & 20171$+$3519 & 13 & 5 \\
16031$-$4856 & 13 & 6     & 17507$-$1122 & 13 & 6   & 20217$+$3330 & 13 & 6 \\
16038$-$5008 & 13 & 6     & 17559$-$2848 & 13 & 6   & 20267$+$2105 & 13 & 6 \\
16055$-$4621 & 13 & 5     & 18006$-$3213 & 13 & 6   & 20365$+$1154 & 13 & 6 \\
16061$-$4555 & 13 & 10    & 18022$-$1432 & 13 & 6   & 20509$+$4212 & 13 & 6 \\
16077$-$5830 & 13 & 6     & 18027$-$2314 & 13 & 6   & 21453$+$5959 & 13 & 6 \\
16109$-$4651 & 13 & 6     & 18046$-$2322 & 13 & 6   & 22525$+$6033 & 13 & 6 \\
16146$-$5257 & 13 & 6     & 18050$-$0518 & 13 & 6   & 16235$-$4832 & 14 & 10 \\
03112$-$5730 & 15 & 1     & 10329$-$3918 & 15 & 2   & 18400$-$0704 & 15 & 13 \\
05418$-$3224 & 15 & 1     & 12226$+$0102 & 15 & 1   & 18551$+$1345 & 15 & 4 \\
05426$+$2040 & 15 & 1     & 12544$+$6615 & 15 & 2   & 19314$-$1629 & 15 & 1 \\
05440$+$4311 & 15 & 1     & 13136$-$4426 & 15 & 1   & 19537$+$2212 & 15 & 1 \\
06183$+$1135 & 15 & 1     & 14232$-$6106 & 15 & 14  & 20014$+$2830 & 15 & 13 \\
06230$-$0930 & 15 & 1     & 15193$-$5656 & 15 & 4   & 15357$-$5239 & 16 & 5 \\
07098$-$2012 & 15 & 1     & 16296$-$4417 & 15 & 13  & 20068$+$3328 & 16 & 7 \\
07538$-$3928 & 15 & 3     & 17044$-$3722 & 15 & 1   & 20286$+$4105 & 16 & 14 \\
08304$-$4313 & 15 & 14    & 17047$-$2848 & 15 & 1   & 21334$+$5039 & 16 & 7 \\
08439$-$2734 & 15 & 1     & 17048$-$1601 & 15 & 1   & 22566$+$5830 & 16 & 0 \\
09521$-$7508 & 15 & 1     & 17544$-$2951 & 15 & 1   & & &\\
10239$-$5818 & 15 & 4     & 18234$-$2206 & 15 & 1   & & &\\
\enddata
\end{deluxetable}

\clearpage


\begin{figure}
\plotone{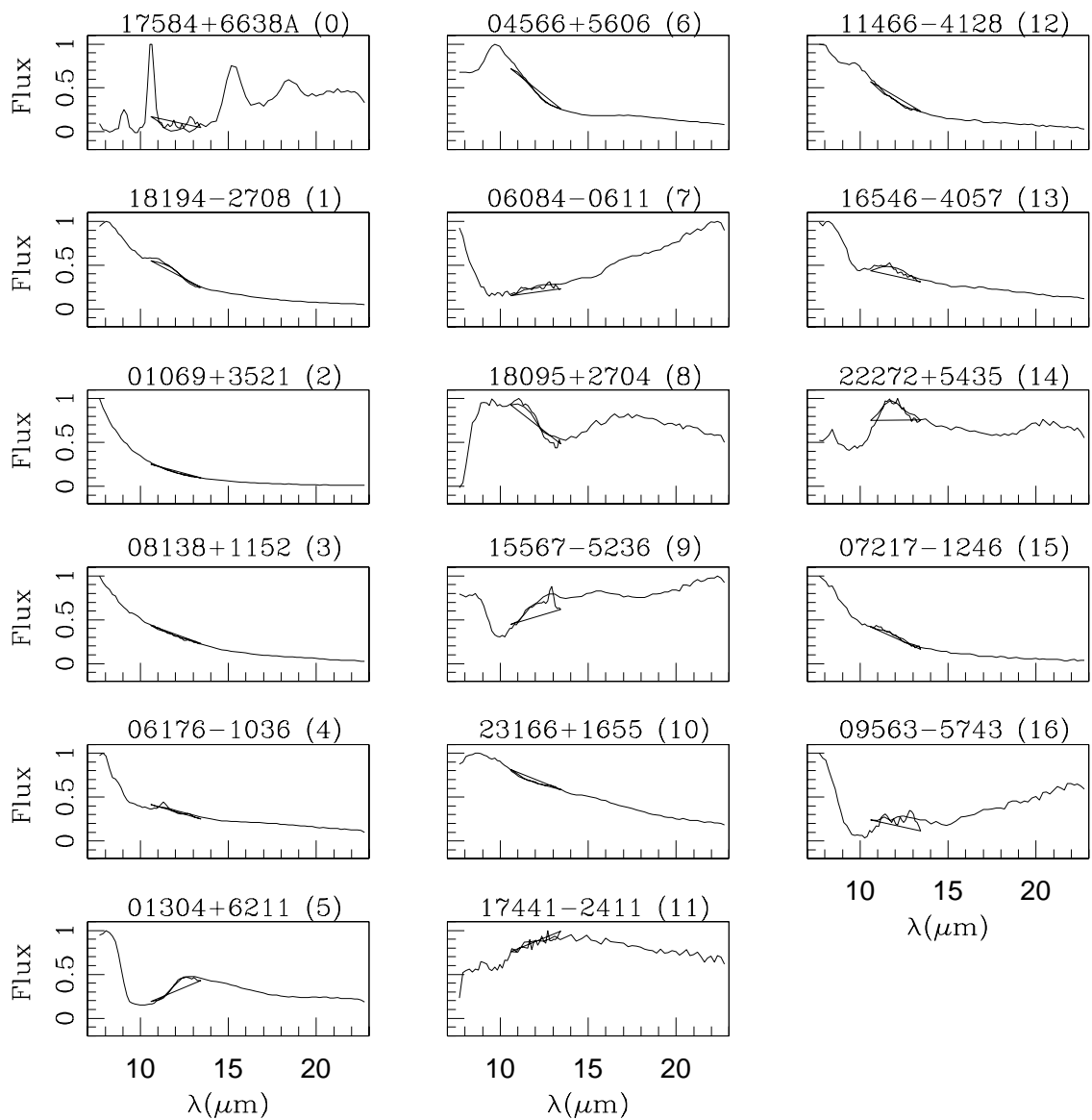}
\caption{The LRS spectra for the 17 training classes. There are ten training
spectra each for the 17 Catalog classes but only one representative spectrum
is shown. For each spectrum the IRAS name
and the Catalog class (in parenthesis) are given above the plot. 
Please note that there exists a crossover region between the "blue"
and the "red" spectra (due to the instrument's settings) ranging between
the 10-14 $\mu m$ and this is seen in these plots.
\label{fig1}}
\end{figure}

\clearpage
\begin{figure}
\figurenum{2}
\plottwo{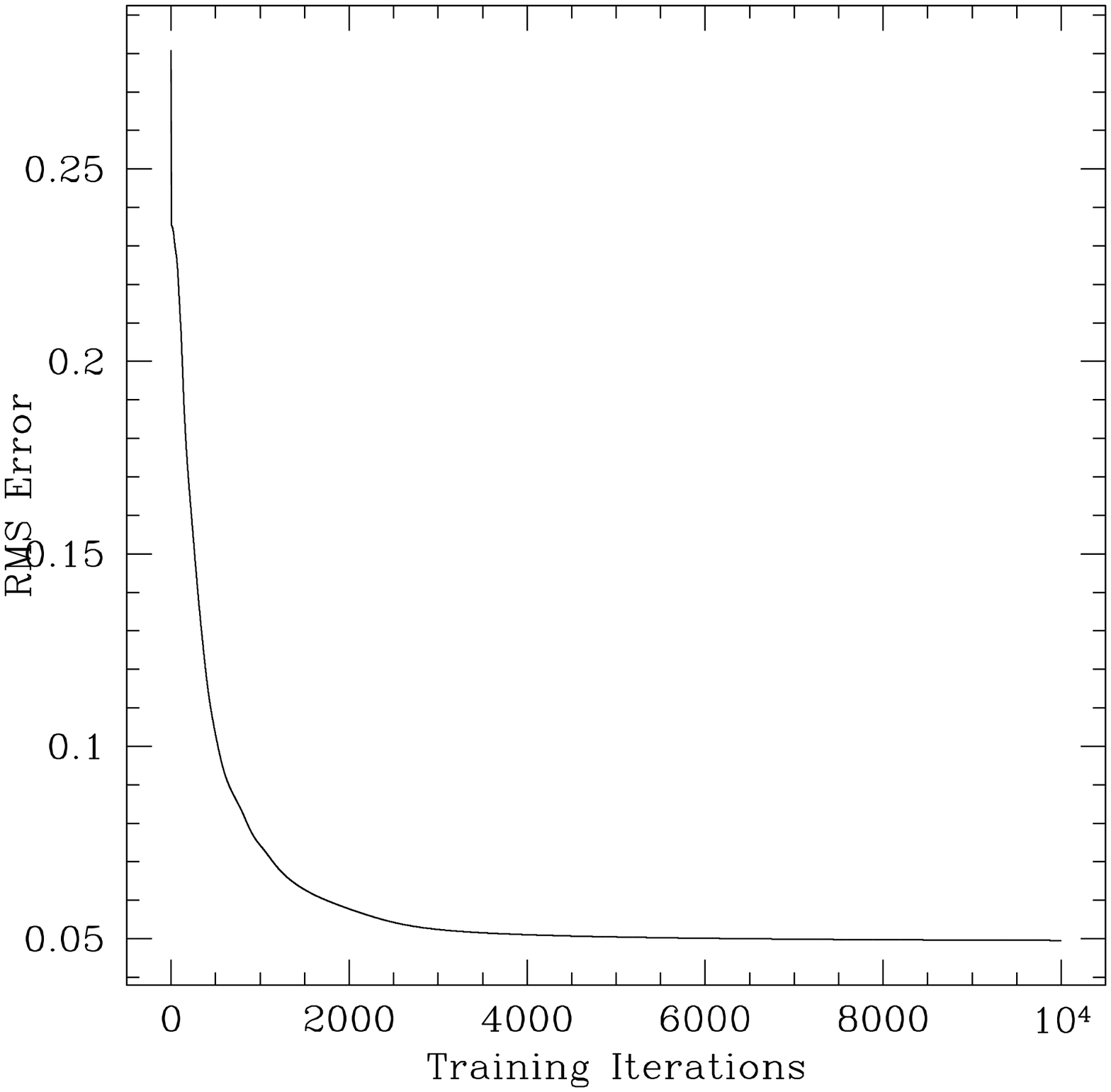}{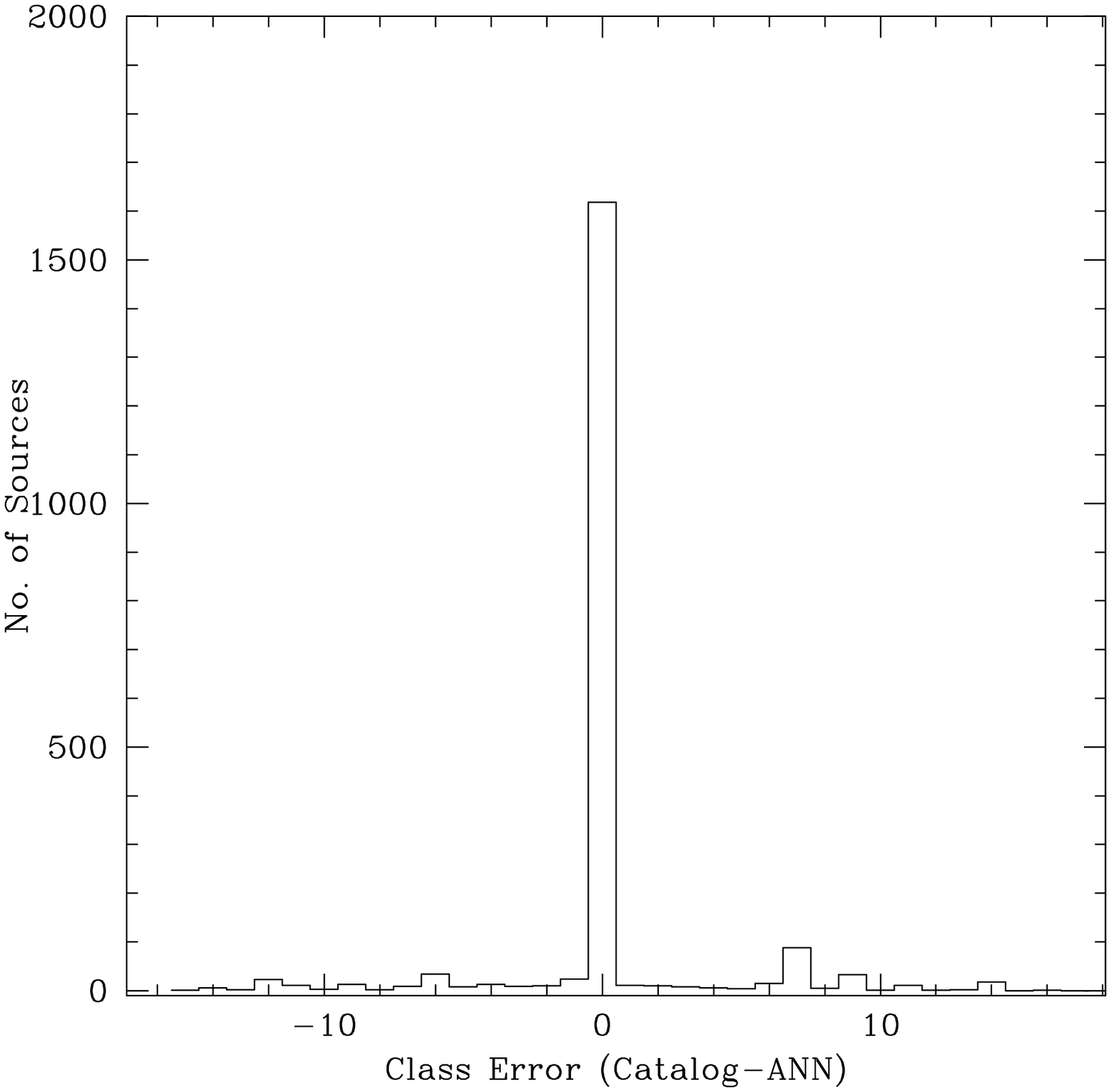}
\caption{ANN learning curve for 10000 iterations (left panel). The
RMS error refers to the RMS difference between the computed
output and the desired output for each iteration. The right panel shows the
histogram of classification accuracy for the 2000 test patterns with 17
training classes.
\label{fig2}}
\end{figure}
\clearpage
\begin{figure}
\figurenum{3}
\plotone{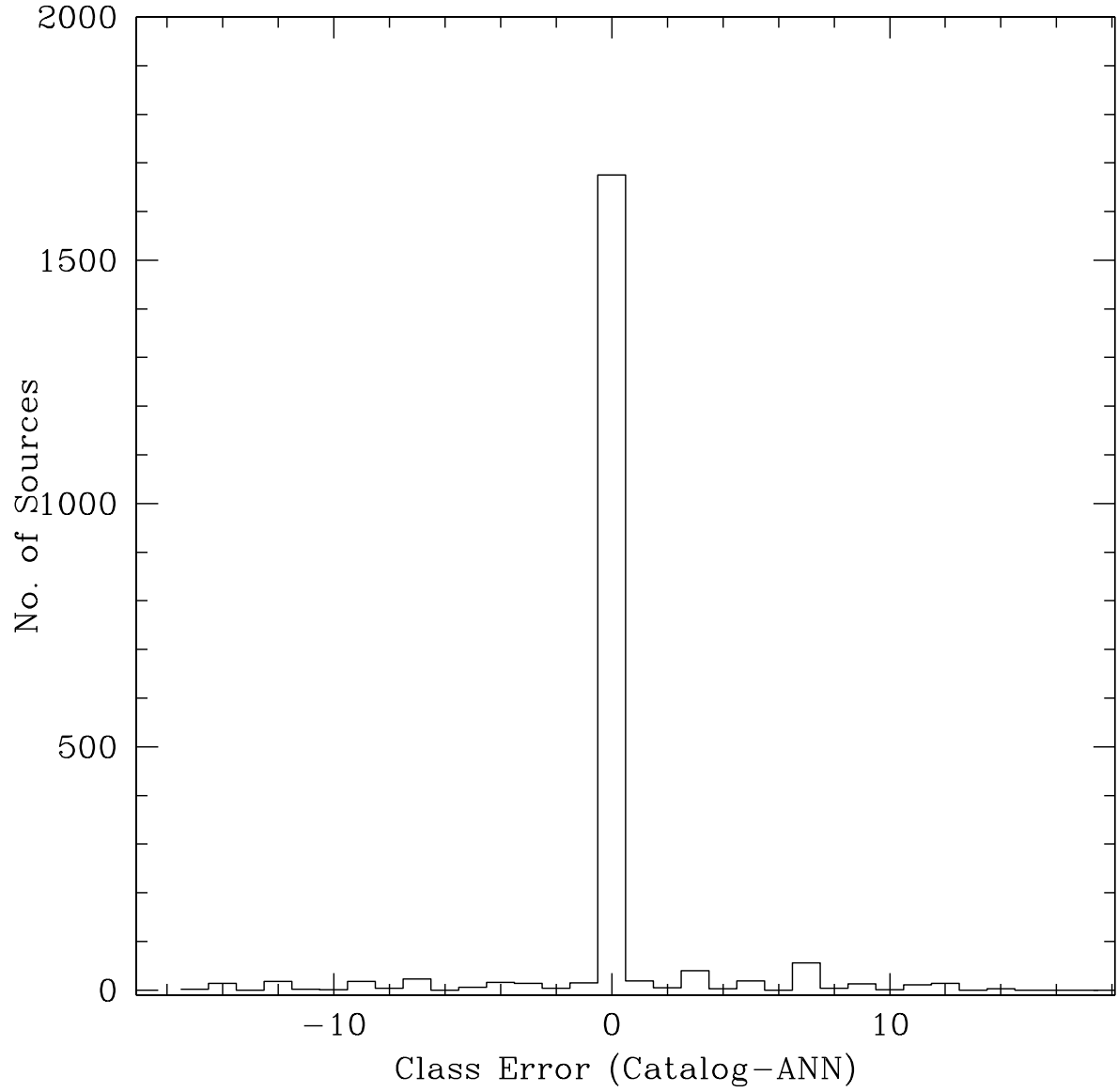}
\caption{The histogram of classification accuracy for the 2000 test
patterns with 16 training classes (class 12 trained as class 6)
\label{fig3}}
\end{figure}
\begin{figure}
\figurenum{4}
\plottwo{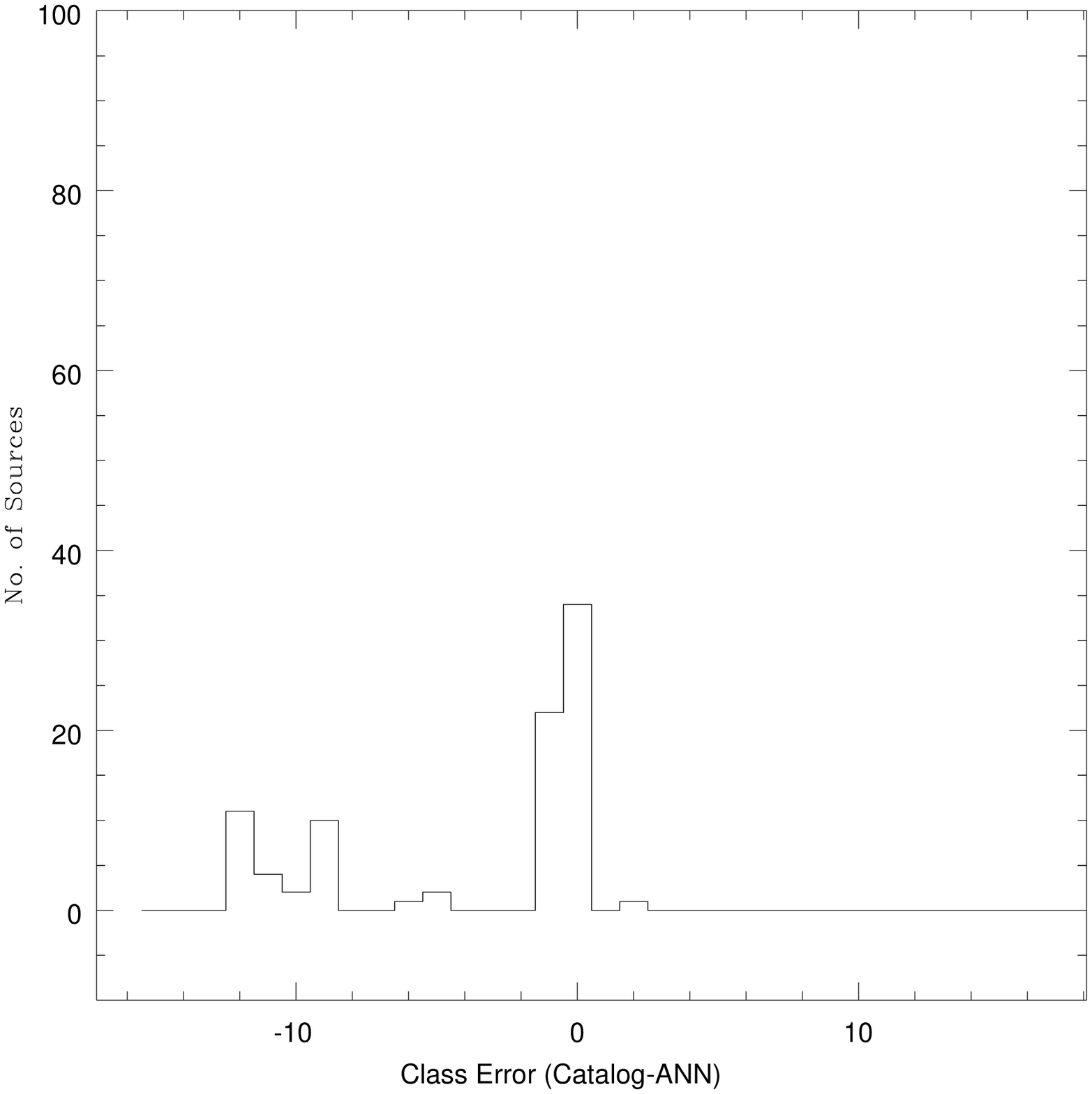}{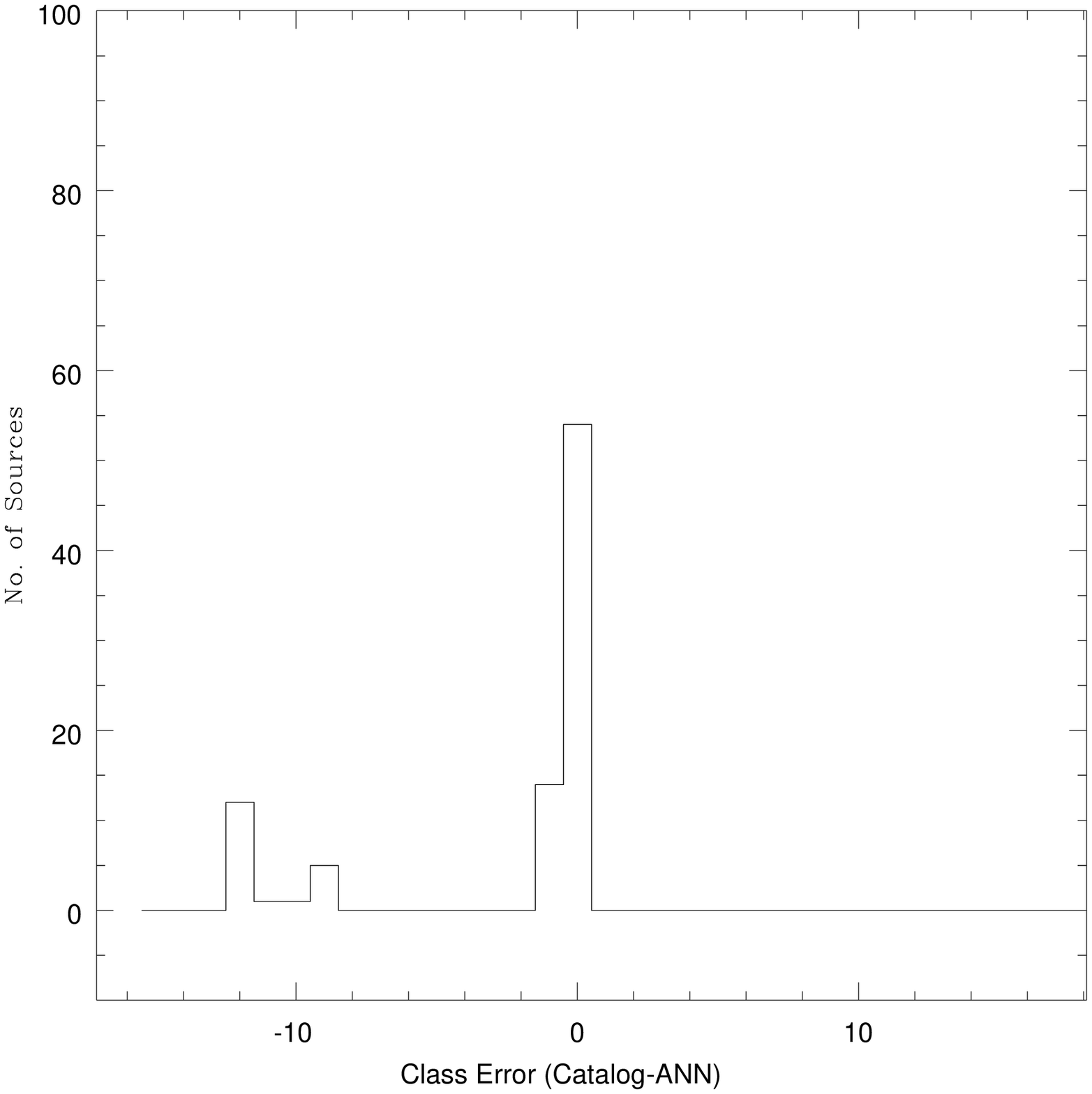}
\caption{Histogram of classification accuracy for the 87 test patterns of class
4 with 17 training classes (left panel) and with combined training
class 6 and class 12 (right panel)
\label{fig4}}
\end{figure}
\clearpage
\begin{figure}
\figurenum{5}
\plottwo{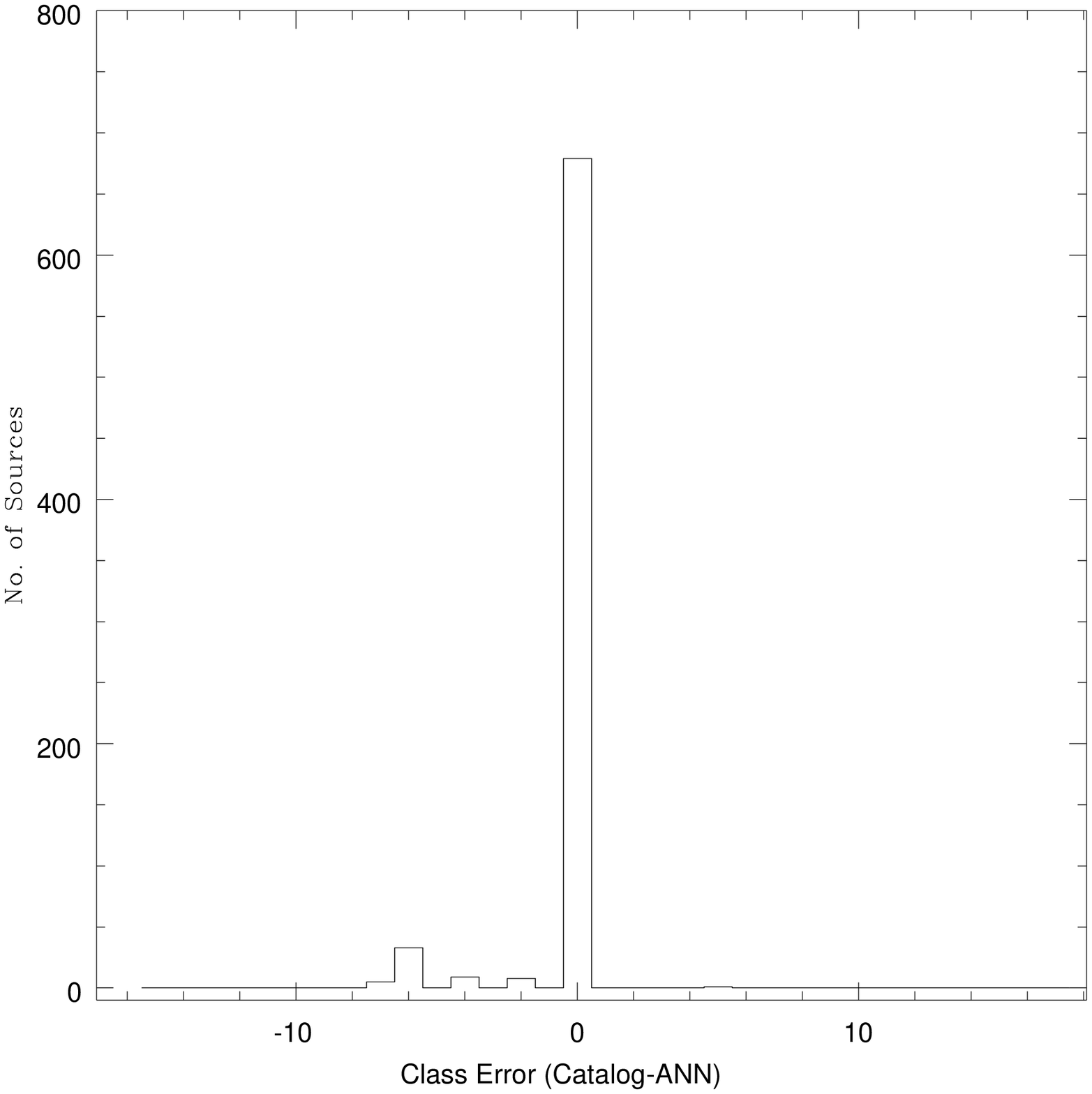}{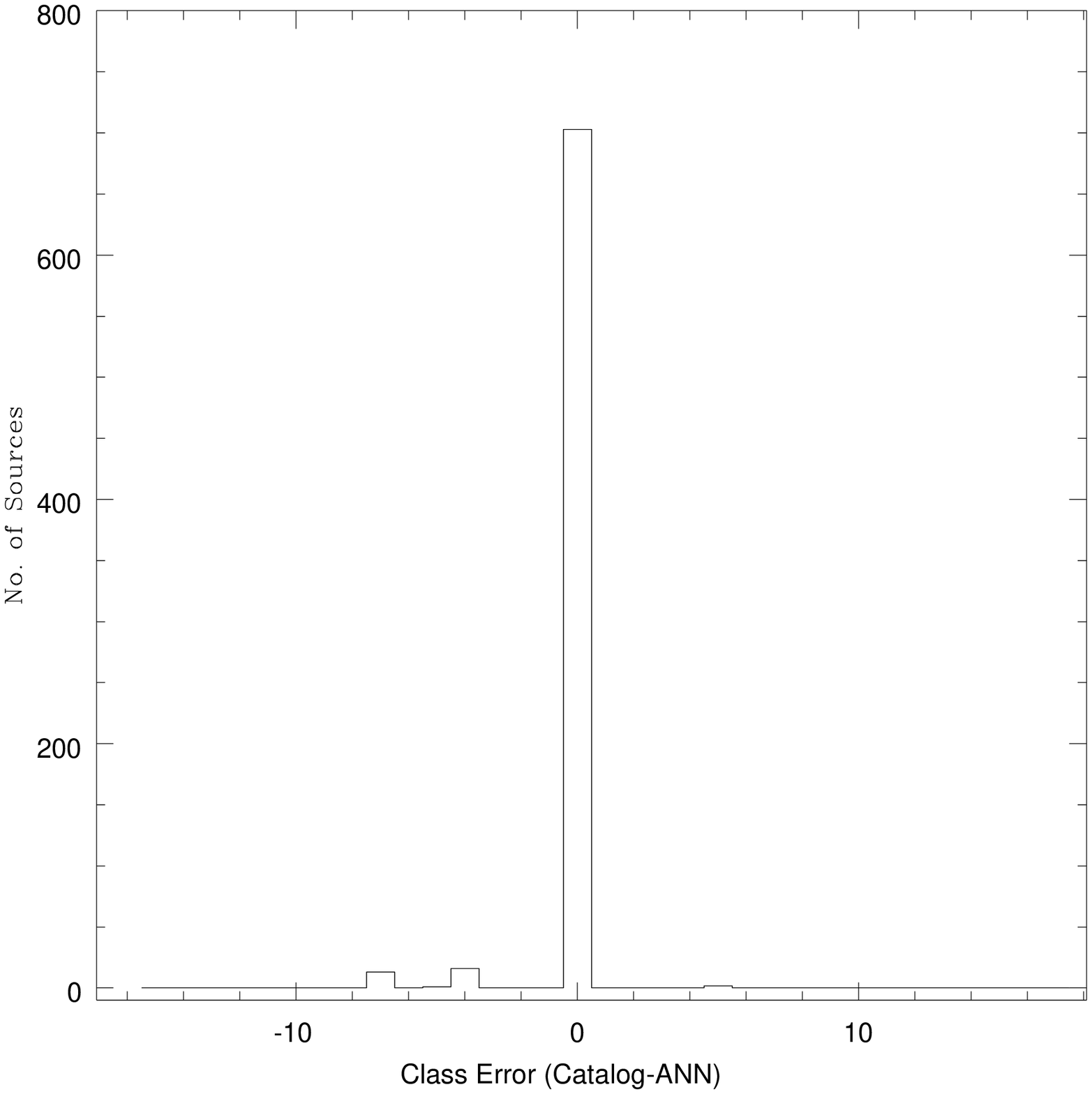}
\caption{Histogram of classification accuracy for the 735 test patterns of class
 6 with 17 training classes (left panel) and with combined training
class 6 and class 12 (right panel)
\label{fig5}}
\end{figure}

\begin{figure}
\figurenum{6}
\plottwo{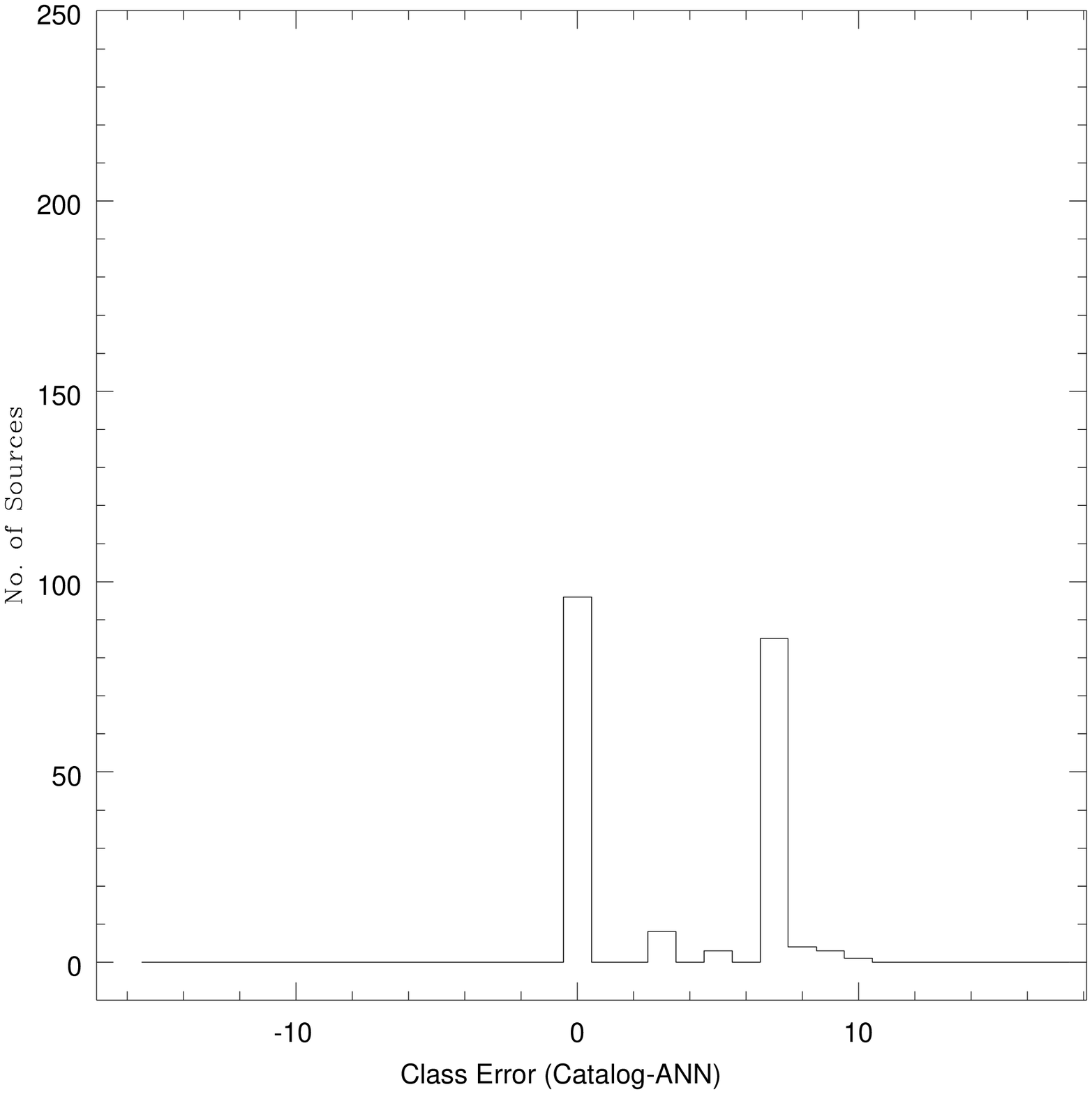}{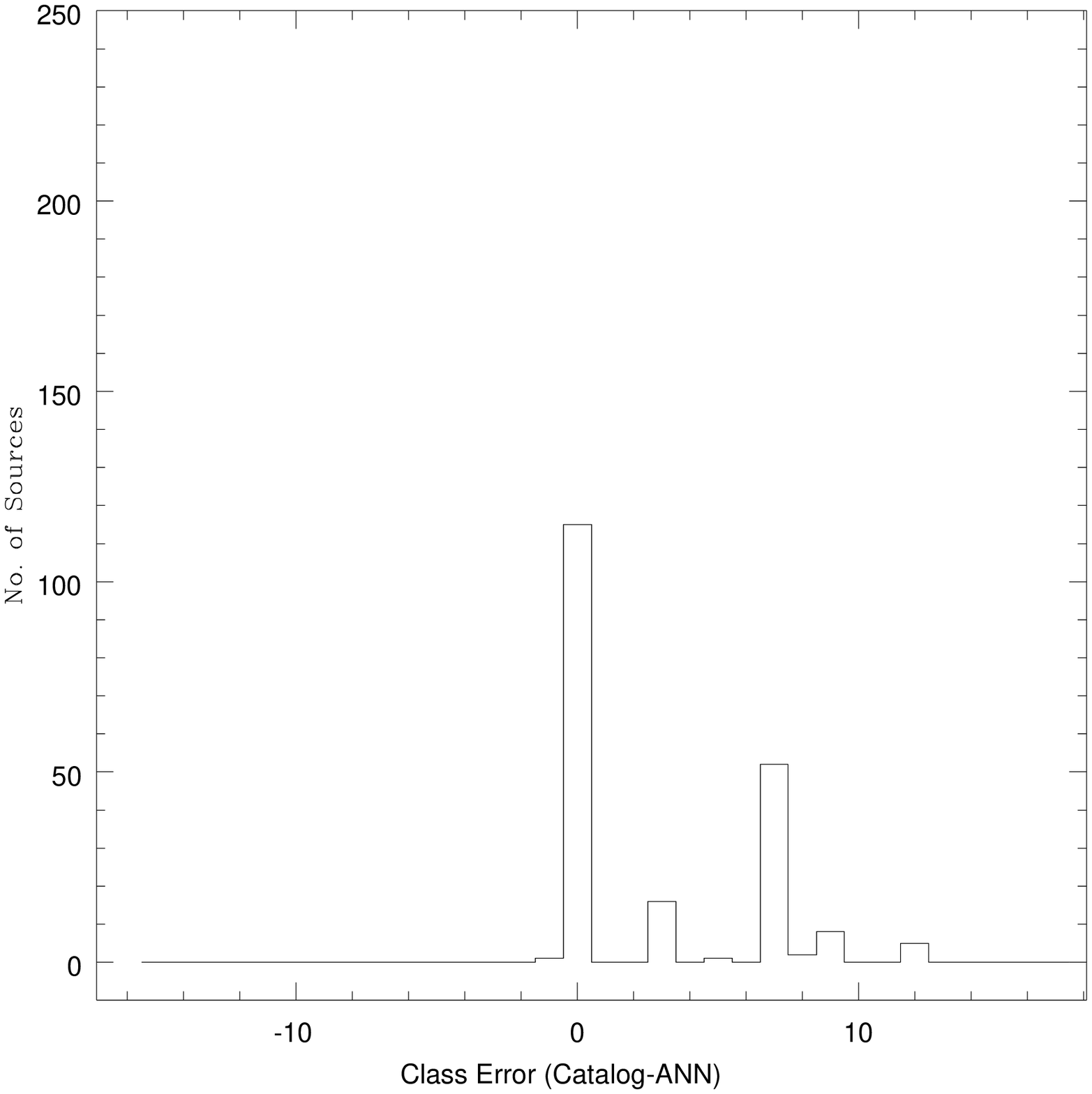}
\caption{Histogram of classification accuracy for the 200 test patterns of
class 13 with 17 training classes (left panel) and with combined
training class 6 and class 12 (right panel)
\label{fig6}}
\end{figure}

\begin{figure}
\figurenum{7}
\plotone{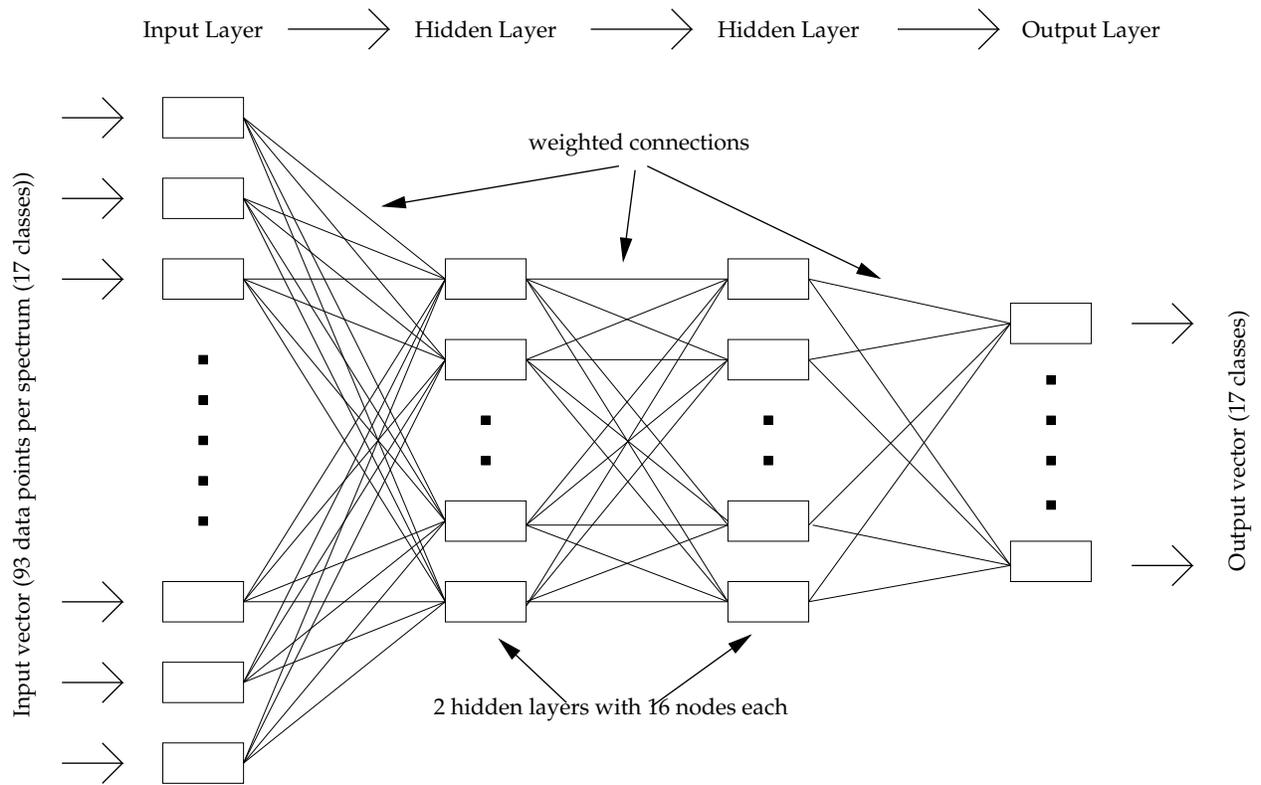}
\caption{Feedforward artificial neural network architecture for the problem
under investigation
\label{fig7}}
\end{figure}
\end{document}